\documentclass{emulateapj}

\usepackage{natbib}
\usepackage{amsmath}
\usepackage{color}

\newcommand{\Ni}{$^{56}$Ni}

\shorttitle{Shock Breakout in Non-Steady Dense Wind}
\shortauthors{Moriya \& Tominaga}

\begin{document}

\title{
Diversity of Luminous Supernovae from Non-Steady Mass Loss
}

\def\ipmu{1}
\def\ut{2}
\def\resceu{3}
\def\konan{4}

\author{
{Takashi J. Moriya}\altaffilmark{\ipmu,\ut,\resceu} and
{Nozomu Tominaga}\altaffilmark{\konan,\ipmu}}

\altaffiltext{\ipmu}{
Institute for the Physics and Mathematics of the Universe,
Todai Institutes for Advanced Study,
University of Tokyo, Kashiwanoha 5-1-5, Kashiwa, Chiba 277-8583, Japan;
takashi.moriya@ipmu.jp
}
\altaffiltext{\ut}{
Department of Astronomy, Graduate School of Science, University of Tokyo,
Hongo 7-3-1, Bunkyo, Tokyo 113-0033, Japan
}
\altaffiltext{\resceu}{
Recearch Center for the Early Universe,
Graduate School of Science, University of Tokyo,
Hongo 7-3-1, Bunkyo, Tokyo 113-0033, Japan
}
\altaffiltext{\konan}{
Department of Physics, Faculty of Science and Engineering, Konan University,
Okamoto 8-9-1, Kobe, Hyogo 658-8501, Japan
}

\begin{abstract}
We show that the diversity in the density slope of the dense wind
due to non-steady mass loss can be one way to 
explain the spectral diversity of Type II luminous supernovae (LSNe).
The interaction of SN ejecta and wind surrounding it is considered
to be a power source
to illuminate LSNe because many LSNe show the wind signature in their
 spectra (Type IIn LSNe).
However, there also exist LSNe without the spectral
features caused by the wind (Type IIL LSNe).
We show that, even if LSNe are illuminated by the interaction,
it is possible that they do not show the narrow spectra from the wind
if we take into account of non-steady mass loss of their progenitors.
When the shock breakout takes place in the dense wind with
the density structure $\rho\propto r^{-w}$, 
the ratio of the diffusion timescale in the optically thick
region of the wind $(t_d)$ and the
shock propagation timescale of the entire wind after
the shock breakout $(t_s)$ strongly depends on $w$.
For the case
$w\hspace{0.3em}\raisebox{0.4ex}{$<$}\hspace{-0.75em}\raisebox{-.7ex}{$\sim$}\hspace{0.3em}1$,
both timescales are comparable ($t_d/t_s\simeq 1$) and
$t_d/t_s$ gets smaller as $w$ gets larger.
For the case $t_d/t_s\simeq 1$,
the shock goes through the entire wind just after
the light curve (LC) peak and narrow spectral lines from
the wind cannot be observed after the LC peak (Type IIL LSNe).
If $t_d/t_s$ is much smaller, the shock wave continues to propagate in
 the wind after the LC peak
and unshocked wind remains (Type IIn LSNe).
This difference can be obtained only through a careful treatment
of the shock breakout condition in a dense wind.
The lack of narrow Lorentzian line profiles in Type IIL LSNe
before the LC peak can also be
explained by the difference in the density slope.
Furthermore, we apply our model to Type IIn LSN 2006gy and
Type IIL LSN 2008es and find that our model is consistent with the observations.
\end{abstract}


\section{Introduction}\label{sec1}
Shock breakout is a phenomenon which is predicted to be observed
when a shock wave emerges from the surface of an exploding star.
Before the shock wave approaches the surface of the star,
the diffusion timescale of photons is much longer than the dynamical
timescale of the shock wave because of the high optical depth of the 
stellar interior and photons cannot go out of the shock wave.
At the stellar surface, the optical depth above the shock wave
suddenly becomes low enough for
photons to diffuse out from the shock wave and photons start
to travel away from the star. This sudden release of photons is
predicted to be observed as a flash of X-rays or ultraviolet (UV) photons
\citep[e.g.,][]{ohyama1963,colgate1974,klein1978,ensman1992,
blinnikov1998,blinnikov2000,matzner1999,calzavara2004,nakar2010,tominaga2011}
and
it is actually observed for several times, e.g., XRO 080109/SN 2008D
\citep[e.g.,][]{soderberg2008} and
SNLS-04D2dc \citep[][]{schawinski2008,gezari2009}.

If the circumstellar wind of the SN progenitor is dense and
optically thick, the shock breakout signal is altered by the wind.
Photons emitted from the shock diffuse in the wind and
the light curve (LC) of the shock breakout becomes broader
\citep[e.g.,][]{falk1977,grasberg1987,moriya2011}.
If the wind is much denser, the shock breakout itself can take place in the wind.
The shock breakout in the dense wind is related to astrophysical
phenomena, e.g.,
PTF 09uj \citep{ofek2010}, XRO 080109 \citep{balberg2011}, and
production of high energy particles \citep{murase2011,katz2011}.
In particular, \citet{chevalier2011} associate luminous supernovae (LSNe)
to the shock breakout in the dense wind.

Many LSNe are believed to be brightened by 
the shock interaction between SN ejecta (or materials released from
stellar surface) and the dense circumstellar wind
\citep[e.g.,][]{smith2007b,woosley2007,vanmarle2010,blinnikov2010}\footnote{
There are other suggestions for energy sources to brighten LSNe: e.g.,
huge amount of \Ni~produced during SN explosions
\citep[e.g.,][]{gal-yam2009,young2010,moriya2010},
newly born magnetars \citep[e.g.,][]{maeda2007,kasen2010,woosley2010},
and optical afterglows of gamma-ray bursts \citep[e.g.,][]{young2005}.}.
This is because many LSNe are spectroscopically classified as Type IIn SNe
which show narrow spectral lines from the wind surrounding SN ejecta\footnote{
See, for example, \citet{schlegel1990,filippenko1997} for the spectral
classification of Type IIn and other SN types.}
\citep[e.g.,][]{smith2010,drake2010,rest2011,chatzopoulos2011,drake2011}.
For example, SN 2006gy shows Lorentzian H Balmer lines
with the full width at half maximum velocity $\simeq
1,000~\mathrm{km~s^{-1}}$ \citep[e.g.,][]{smith2010}
which are suggested to originate from the dense wind
\citep[e.g.,][]{chugai2001,chugai2004,dessart2009}.
In addition, SN 2006gy shows narrow P-Cygni profiles from
a $\simeq 100~\mathrm{km~s^{-1}}$ outflow which is also presumed to stem
from the wind surrounding the SN ejecta \citep[e.g.,][]{smith2010}.
However, there commonly exist LSNe without narrow spectral lines
from the circumstellar wind
\citep[e.g.,][]{quimby2007,miller2009,gezari2009,pastorello2010,quimby2011,chomiuk2011}.
Based on the LC shapes of LSNe, \citet{chevalier2011} show that,
if the shock breakout takes place inside the dense wind,
the SN is observed as SN 2006gy-like Type IIn LSNe and, if it takes place at the surface,
the SN is observed as SN 2010gx-like non-Type IIn LSNe.

In the previous works on the shock breakout in a dense wind,
the density slope of the wind has been assumed to be $\simeq-2$
which is a consequence of the steady mass loss of the progenitor.
However, non-steady mass loss is actually observed
in the massive stars which are suggested to be possible progenitors of
LSNe, e.g., $\eta$ Carinae \citep[e.g.,][]{davidson1997,gal-yam2009b}.
In addition, historical X-ray SN observations are suggested to
indicate that mass loss from progenitors of Type IIn SNe
is inconsistent with the steady mass loss \citep{dwarkadas2011b}.
Here, we investigate the influence of non-steady mass loss on the shock breakout
in the dense wind and show that the diversity in the wind density slope
caused by the non-steady mass loss
can explain why some Type II LSNe show narrow spectral
components from the wind (Type IIn) while others do not (Type IIL)
even if both LSNe are illuminated by the shock interaction.
Our model is presented in Section \ref{sec2} and it is applied to LSNe in Section \ref{sec3}.
We discuss our results in Section \ref{sec4}. Conclusions are summarized in Section \ref{sec5}.

\section{Effect of Dense Non-Steady Wind}\label{sec2}
\subsection{Wind Configuration}\label{sec2.1}
We consider a spherically symmetric dense wind extending
from $r=R_i$ to $r=R_o$ where $r$ is radius.
The density $\rho$ is assumed to follow $\rho (r)=Dr^{-w}$ with a constant $D$.
The opacity $\kappa$ of the wind is assumed to be constant throughout this paper.
Although our model is applicable to any uniform compositions with a
constant $\kappa$, we mainly consider H-rich winds throughout this paper.
We assume that a SN explosion has occurred in the dense wind and
that the shock breakout occurs in the wind
\citep[e.g.,][see Section \ref{sec1}]{chevalier2011}.
The radius of the forward shock at the time of the shock breakout is set to $r=xR_o$.
Note that the entire wind above the progenitor star 
is optically thick because of the assumption
that the shock breakout occurs in the dense wind.
We also introduce a radius $y_\tau R_o$ where the optical depth
evaluated from the wind surface
becomes $\tau$, i.e.,
\begin{equation}
\tau=\int^{Ro}_{y_\tau R_o}\kappa\rho dr.\label{ydef}
\end{equation}

\begin{figure*}
\begin{center}
\includegraphics[width=\columnwidth]{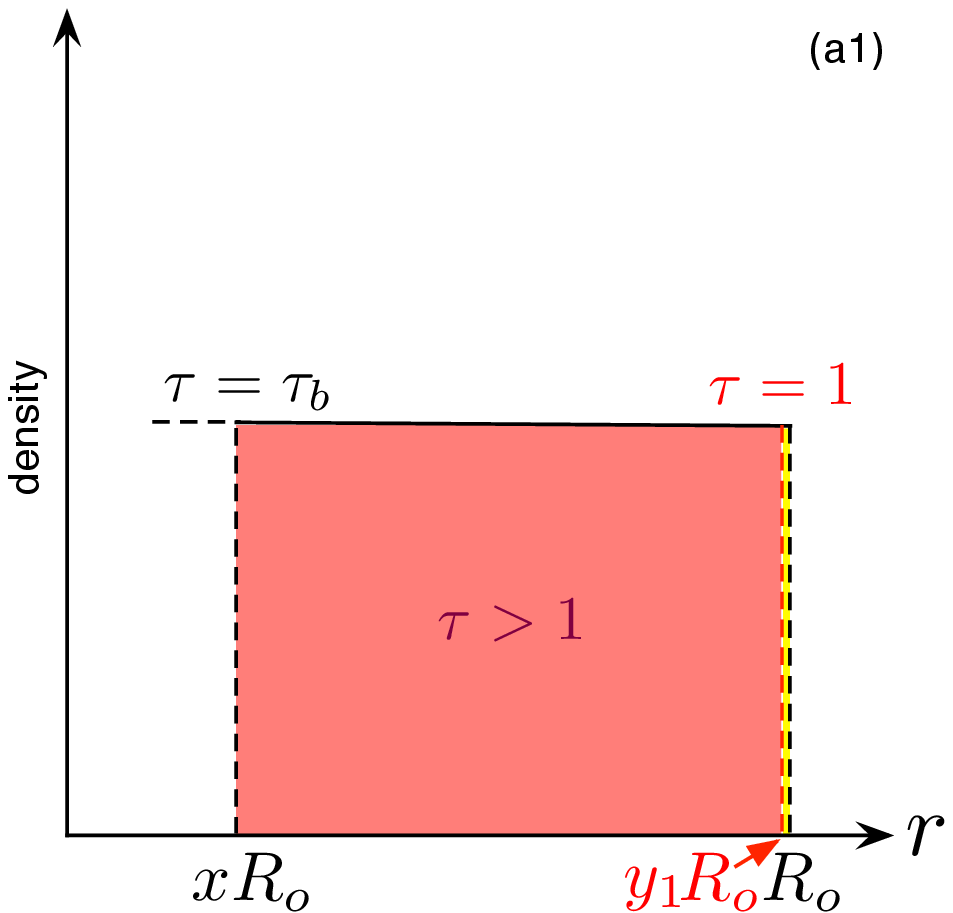}
\includegraphics[width=\columnwidth]{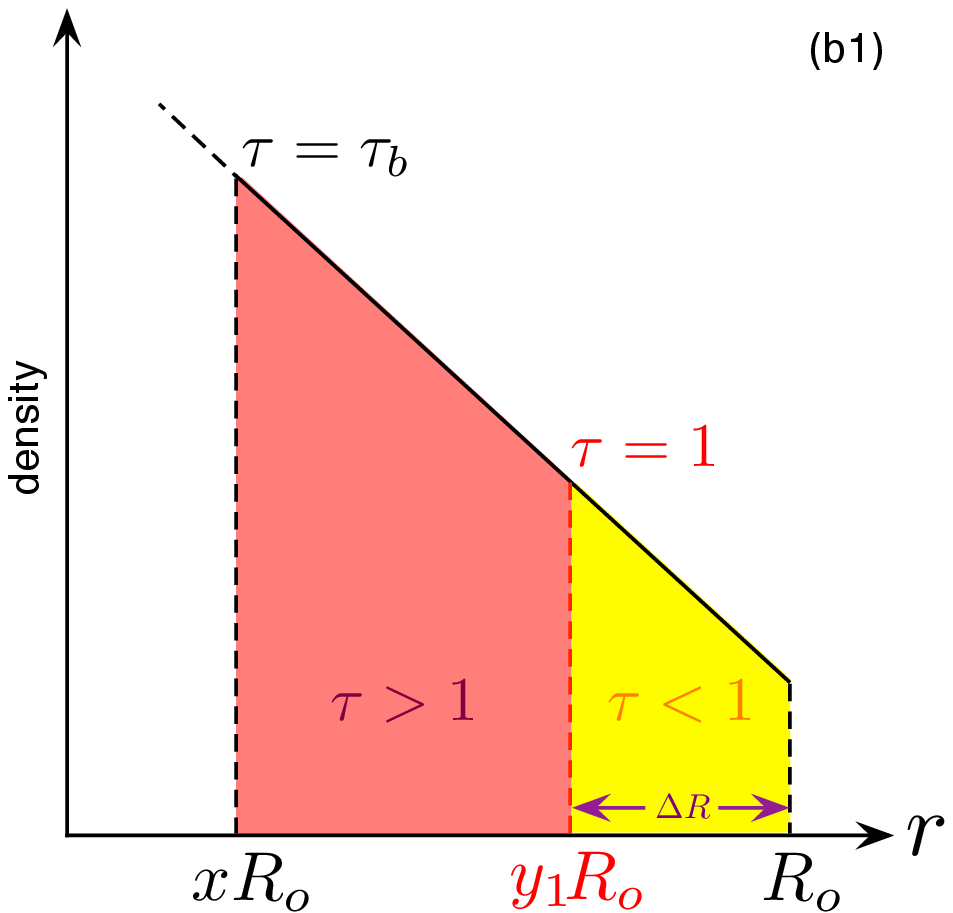}\\
\includegraphics[width=\columnwidth]{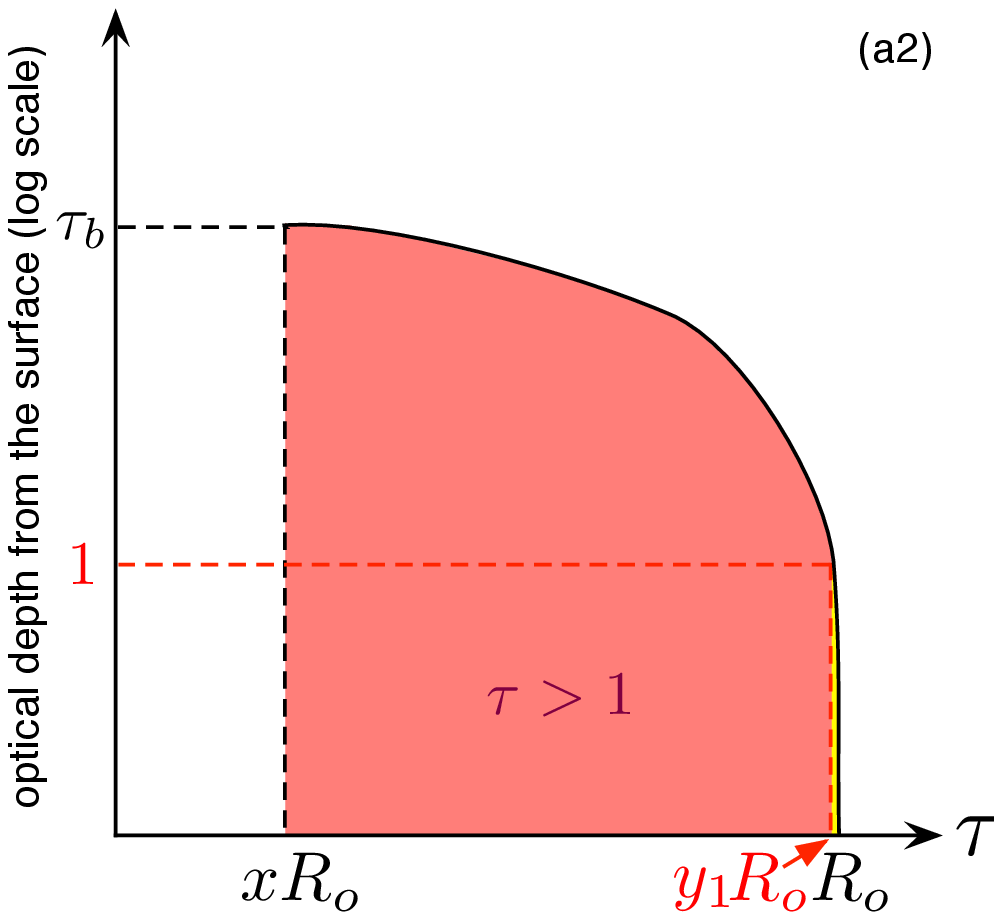}
\includegraphics[width=\columnwidth]{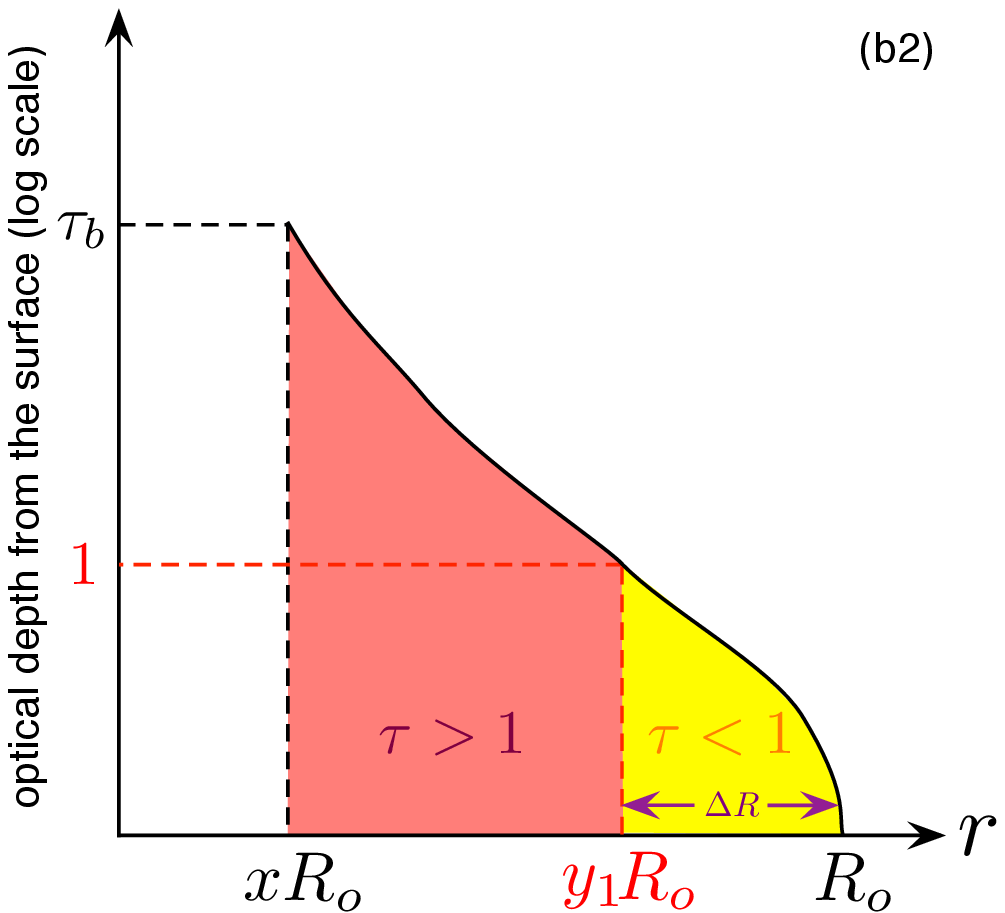}
\caption{
Illustration of the density and the optical depth distributions
for the winds with different density slopes:
(a) a flat density slope and (b) a steep density slope.
The corresponding (1) density structures and (2) optical depth distributions
are shown.
The shock breakout is assumed to occur at the same $r=xR_o$ with
$x<1$ where the opacity $\tau$ from $R_o$ becomes $\tau_b$.
All the opacities shown in the figure is evaluated from $R_o$.
Large $\tau<1$ region appears in the outer part of the wind in (b).
See Figure \ref{fig4} for the concrete examples of the distribution.
}\label{fig1}
\end{center}
\end{figure*}

\subsection{Shock Breakout Condition in Non-Steady Dense Wind}
Shock breakout in a wind had been simply
defined to occur when the diffusion timescale of the entire
wind is comparable to the shock propagation timescale of the entire wind, i.e.,
\begin{equation}
\int^{R_o}_{xR_o}\kappa \rho dr\simeq\frac{c}{v_s} \label{tsb}
\end{equation}
\citep[see, e.g.,][for the details of the shock breakout condition]
{weaver1976,nakar2010}.

\begin{figure}
\begin{center}
\includegraphics[width=\columnwidth]{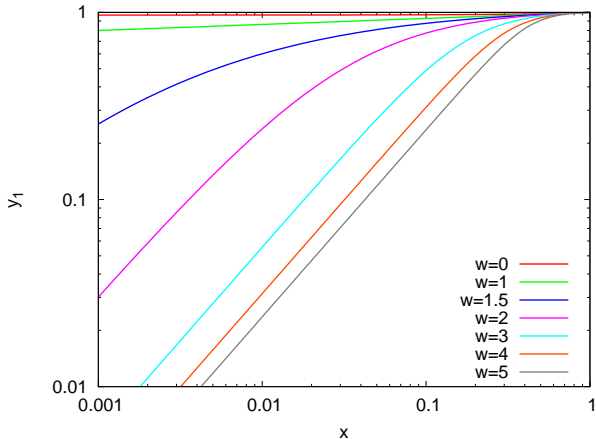}
\caption{
Location of the last scattering surface in the wind.
The region $r>y_1R_o$ is optically thin and photons do not diffuse in
the region.
}\label{fig2}
\end{center}
\end{figure}

However, the wind could contain a large optically thin
region even if the entire wind
is optically thick.
Figure \ref{fig1} is a simplified illustration of the effect of $w$
in the dense wind.
Two dense winds with the different $w$ but the same $R_o$
and $xR_o$ are compared in the figure:
(a) a wind with a constant density
and
(b) a wind with a steep density gradient.
In both cases, the shock breakout is assumed to occur in the wind
at the same radius 
$r=xR_o$ $(x<1)$ with the same forward shock velocity $v_s$
and thus the optical depth between
$xR_o$ and $R_o$ is exactly the same in both cases $(\tau_b)$.
One of the important differences in the two winds are in the radius
of the last scattering surface $y_1R_o$, where $\tau=1$
(see Figure \ref{fig2} in the following for the value of $y_1$).
Even if the entire wind is optically thick in both cases,
the region where $\tau$ becomes larger than 1 $(r<y_1R_o)$
is more concentrated to the central region
and the wind contains larger
optically thin region outside whose size is $\Delta R=R_o-y_1R_o$
in the case of large $w$.

If the wind contains an extended optically thin region,
Equation (\ref{tsb}) is no longer an appropriate condition
for the shock breakout because it postulates that the entire wind
at $r\leq R_o$ is optically thick enough for photons to be diffusive.
The shock breakout condition should be evaluated only at the optically
thick region where photons are diffusive.
Thus, the shock breakout should occur
when the diffusion timescale of the optically
thick region of the wind is comparable
to the shock propagation timescale of the region.
Hence, the shock breakout condition should be set as
\begin{equation}
\tau_x\equiv\int^{y_1 R_o}_{xR_o}\kappa \rho dr\simeq\frac{c}{v_s},\label{breakout}
\end{equation}
where $c$ is the speed of light and $\tau_x=\tau_b-1$.
In Equation (\ref{breakout}),
we presume that photons diffuse in the region
where the optical depth evaluated from the wind surface
exceeds $1$ for simplicity.
I.e., photons are assumed to diffuse at
$R_i<r<y_1R_o$ and freely stream at $y_1R_o<r<R_o$.
In this sense, Equation (\ref{breakout}) may also be interpreted
as Equation (\ref{tsb}) combined with a kind of flux limited diffusion approximation.
For the case of the shock breakout at the surface of the wind $(x\simeq 1)$,
the conditions of Equations (\ref{tsb}) and (\ref{breakout})
are similar.

Since the observations of Type II LSNe display a constant shock velocity
until some time in the declining phase of their LCs
\citep[e.g.,][]{smith2010},
we assume that
$v_s$ is constant.
This assumption can be wrong if the shocked wind mass
becomes comparable to the SN ejecta mass.
Equations (\ref{ydef}) and (\ref{breakout}) lead us to
\begin{eqnarray}
D&\simeq&\left\{ \begin{array}{lll}
\frac{\left(1-w\right)\left(1+c/v_s\right)}{\kappa \left(1-x^{1-w}\right)R_o^{1-w}} && (w\neq1),\\ \\
\frac{1+c/v_s}{\kappa (-\ln x)} && (w=1).
\end{array} \right. \label{d}
\end{eqnarray}
In addition, using Equation (\ref{d}), we can express $y_\tau$ as 
\begin{eqnarray}
y_\tau&\simeq&\left\{ \begin{array}{lll}
\left(\frac{c/v_s+\tau x^{1-w}}{c/v_s+\tau}\right)^{\frac{1}{1-w}}&& (w\neq1),\\ \\
x^{\frac{\tau}{c/v_s+\tau}} && (w=1).
\end{array} \right. \label{ytau}
\end{eqnarray}
Figure \ref{fig2} shows a plot of $y_1$ as a function of $x$ for several
$w$.
Larger $w$ leads to smaller $y_1$ for a given $x$,
as is discussed qualitatively above.

Finally, the wind mass $M_\mathrm{wind}$ is defined as
\begin{equation}
M_\mathrm{wind}=\int_{R_i}^{R_o}4\pi r^2\rho dr.
\end{equation}
Using Equation (\ref{d}), we get
\begin{eqnarray}
M_\mathrm{wind}\simeq
\left\{ \begin{array}{lll}
\frac{4\pi (1-w)(1+c/v_s)(R_o^{3-w}-R_i^{3-w})}{\kappa (3-w)(1-x^{1-w})R_o^{1-w}}&& (w\neq1,3),\\ \\
\frac{2\pi (1+c/v_s)(R_o^{2}-R_i^{2})}{\kappa (-\ln x)} && (w=1), \\ \\
\frac{8\pi (1+c/v_s)\ln (R_o/R_i)}{\kappa (x^{-2}-1) R_o^{-2}} && (w=3).
\end{array} \right. \label{Mwind}
\end{eqnarray}

\subsection{Timescales of Photon Diffusion and Shock Propagation in Dense Wind}
Here, we estimate the timescale of photon diffusion $(t_d)$
which characterizes the LCs of LSNe and the timescale of
the shock propagation $(t_s)$ which represents the timescale for the
forward shock to go through the entire wind.
$t_d$ corresponds to the timescale for the LC to reach the peak
\citep[e.g.,][]{arnett1980,arnett1982}.
$t_d$ can be expressed as
\begin{eqnarray}
t_d&\simeq&\frac{\tau_x (R_o-xR_o-\Delta R)}{c},\\
&=&\frac{\tau_x (y_1R_o-xR_o)}{c},\\
&=&\left\{ \begin{array}{lll}
\frac{R_o}{v_s}\left[\left(\frac{c/v_s+ x^{1-w}}{c/v_s+1}\right)^{\frac{1}{1-w}}-x\right]&& (w\neq1),\\ \\
\frac{R_o}{v_s}\left(x^{\frac{1}{1+c/v_s}}-x\right) && (w=1).
\end{array} \right. \label{td}
\end{eqnarray}
$t_s$ is defined as the time required for the forward shock to go through the
entire wind including the optically thin region after the shock breakout,
\begin{equation}
t_s=\frac{R_o-xR_o}{v_s}.\label{ts}
\end{equation}
Hence, we can derive the ratio of the two timescales:
\begin{eqnarray}
\frac{t_d}{t_s}&\simeq&\left\{ \begin{array}{lll}
\frac{1}{1-x}\left[\left(\frac{c/v_s+ x^{1-w}}{c/v_s+1}\right)^{\frac{1}{1-w}}-x\right]&& (w\neq1),\\ \\
\frac{1}{1-x}\left(x^{\frac{1}{1+c/v_s}}-x\right) && (w=1).
\end{array} \right. \label{td/ts}
\end{eqnarray}

\begin{figure}
\begin{center}
\includegraphics[width=\columnwidth]{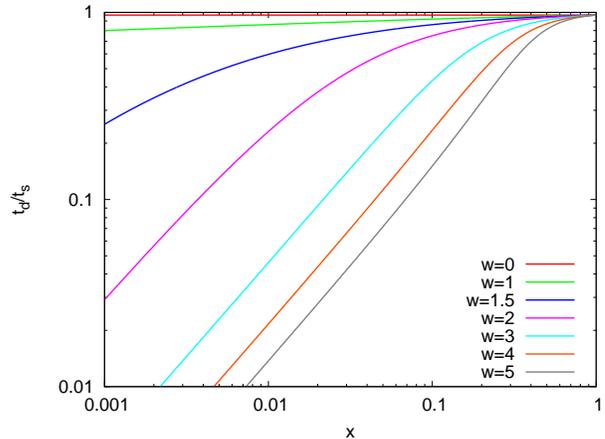}
\caption{
Ratio of the diffusion timescale ($t_d$)
and the shock propagation timescale ($t_s$)
as a function of the location of the shock breakout ($x$).
}\label{fig3}
\end{center}
\end{figure}

Figure \ref{fig3} shows the ratio as a function of $x$ with several $w$,
in which $v_s$ is set to $10,000~\mathrm{km~s^{-1}}$
corresponding to the observational value of SN 2008es (see Section \ref{sec3.2}).
Every line reaches $t_d/t_s\simeq c/\left(v_s+c\right)= 0.97$ at $x\simeq1$.
This corresponds to the case of the shock breakout at the surface of the
wind \citep[e.g.,][]{chevalier2011}.

When the shock breakout occurs inside the wind $(x<1)$,
the ratio $t_d/t_s$ varies depending on the density slope of the wind.
For a given $x$, $t_d/t_s$ gets smaller as 
the density slope of the wind gets steeper (i.e., larger $w$).
This is because the last scattering surface of the wind locates farther
away from the surface as
the density slope gets steeper, i.e., $y_1$ gets smaller as $w$ gets larger
for a given $x$ (see Figure \ref{fig2} for values of $y_1$).
In other words, the optically thin region in the wind is spatially
larger for the wind with steeper density gradient.
This is true even if the wind radius is the same.
We note again that, when the shock breakout occurs in the wind,
the forward shock reaches the last scattering surface
$(r=y_1R_o)$ with the diffusion
timescale $t_d$, i.e., the shock locates at $r=y_1R_o$ at the LC peak.

\subsection{Observational Features and Diversities of Interaction-Powered LSNe}
In this section, we present consequences from the variation in $t_d/t_s$
which results from the variation in $w$.
We show that the two kinds of Type II LSNe, i.e., Type IIn LSNe
\citep[e.g., SN 2006gy,][]{smith2010} and 
Type IIL LSNe  \citep[e.g., SN 2008es,][]{miller2009,gezari2009},
are naturally expected from the shock breakout in the dense H-rich wind
with the different density slope.

If we assume that both types of LSNe are illuminated by the interaction of
the H-rich dense wind and SN ejecta and that the wind is so dense that the
shock breakout occurs in the wind,
the spectral evolution of LSNe is determined by $t_d/t_s$.

If $t_d/t_s\simeq 1$, the shock wave reaches
the surface of the dense wind soon after
the LC has reached the peak with the timescale $t_d$.
Since the entire wind is shocked just after the LC peak,
no signature in spectra from the wind
is observable after the LC peak.
On the other hand, if $t_d/t_s< 1$, the shock wave continues to
propagate in the optically thin region of the wind
even after the LC peak. As there remain unshocked materials in the wind
even after the LC peak, we expect to see narrow P-Cygni profiles from the
unshocked wind even after the LC peak.

To sum up, 
Type IIL LSNe can come from the dense wind with $t_d/t_s \simeq 1$ while
Type IIn LSNe can result from the dense wind with $t_d/t_s < 1$.
The ratio of the two timescales is determined by $w$.

The narrow Lorentzian line profiles which are suggested to be caused by
the dense wind \citep[e.g.,][]{chugai2001,dessart2009} can
appear before the LC peak depending on the optical depth of the wind.
If we apply the model of \citet[][]{chugai2001},
the ratio $U$ of the unscattered H$\alpha$ line flux
to the total H$\alpha$ line flux of the Lorentzian profile is 
\begin{equation}
U\simeq\frac{1-e^{-\tau}}{\tau}.
\end{equation}
For the case of flat density slopes, $\tau$ remains 
too high until the forward shock reaches the surface (Figure \ref{fig1})
and $U$ is expected to be very small for a long time before the LC peak.
On the other hand, if the density decline is steep,
the optical depth decreases gradually with time and
the suitable optical depth for the appearance of Lorentzian profiles
would be realized for a long while.
Therefore,
Lorentzian lines are expected to be observed well before the LC peak
for Type IIn LSNe.

The LC evolution of Type II LSNe is also consistent with our models.
In our models for both Type IIn and Type IIL LSNe,
the forward shock stays in the dense wind until the LC peak.
As the wind with $\tau>1$ is shocked with the timescale of $t_d$,
the dense wind adiabatically cools down after the LC peak.
Thus, the LCs of Type II LSNe are supposed to
follow the shell-shocked diffusion model
presented by \citet{smith2007b}.
The shell-shocked diffusion model is based
on the adiabatic cooling of the shocked dense wind,
which is basically the same as the LC model suggested for Type II SNe by
\citet{arnett1980}, and 
the model had been already shown to be consistent
with the declining phase of the LC of SN 2006gy \citep{smith2007b}.
However, it should be noted that the model is too simplified and
many effects which cannot be treated by the formulation of
\citet{arnett1980} are ignored in the model.
For example, the model assumes a constant opacity and it ignores
the presence of a recombination wave which is supposed to
be created in the diffusing shocked shell.
Thus, we cannot confirm that our model is consistent with
the LCs of Type II LSNe
just by the comparison with the shell-shocked diffusion model
and numerical LC modeling is required to
see if our models are consistent with Type II LSN LCs.

\section{Comparison with Observations}\label{sec3}
In the previous section, we have shown that, if the shock breakout occurs
inside a dense wind $(x<1)$, the ratio of the 
timescale of the photon diffusion to
that of the shock propagation in the wind
depends on the wind density slope and thus the different density slope
can result in two kinds of Type II LSNe, i.e., Type IIn and Type IIL LSNe.
As an example, we apply our model to two LSNe:
Type IIn SN 2006gy and Type IIL SN 2008es.
If we look into Type IIn LSN 2006gy and Type IIL LSN 2008es,
one important difference is the existence of narrow P-Cygni profiles
in the spectra of SN 2006gy after the LC peak.
Based on the observational feature, we can guess that
Type IIL LSN 2008es came from the dense wind with $t_d/t_s \simeq 1$ while
Type IIn LSN 2006gy resulted from the dense wind with $t_d/t_s < 1$.
We apply those models to the two LSNe.
In this section, $\kappa$ is set to $0.34~\mathrm{cm^{2}~g^{-1}}$.

\subsection{Type IIn LSNe (SN 2006gy)}
SN 2006gy is extensively studied by, e.g.,
\citet[][]{ofek2007,smith2007a,smith2007b,woosley2007,smith2008,
agnoletto2009,kawabata2009,smith2010,miller2010}.
It is classified as Type IIn and the luminosity reaches $\sim -22$ mag in the $R$ band \citep{smith2007a}.
The detailed spectral evolution is summarized in \citet{smith2010}.
The narrow P-Cygni H$\alpha$ lines with the absorption minimum of $\simeq 100~\mathrm{km~s^{-1}}$
are considered to come from the wind surrounding the progenitor of SN 2006gy.
As SN 2006gy shows narrow P-Cygni profiles after the maximum luminosity,
an unshocked wind is supposed to remain after the maximum.
Thus, models with $t_d/t_s <1 $ and $y_1<1$ are preferred.
Based on the observations of \citet{smith2010}, we adopt the following parameters:
\begin{eqnarray}
v_s&\simeq&5,200~\mathrm{km~s^{-1}}, \\
t_d&\simeq&70~\mathrm{days}.
\end{eqnarray}
$v_s$ is constrained by the evolution of the blackbody radius
and $t_d$ is obtained from the rising time of the LC.
As the narrow H$\alpha$ P-Cygni profile is detected
at 179 days\footnote{Days since the explosion. The explosion date is set
to be the same as in \citet{smith2010}.}
and disappears at 209 days \citep[][]{smith2010},
we presume that the forward shock has gone through the entire wind between 179 days and 209 days.
We simply take the central date (194 days) as the time when the forward
shock has gone through the entire wind, i.e., $t_s\simeq194$ days.
With $t_d$, $t_s$, and $v_s$, we can estimate $x$ and $R_o$ for a given $w$
from Equations (\ref{td}) and (\ref{ts}).

\begin{figure*}
\begin{center}
\includegraphics[width=\columnwidth]{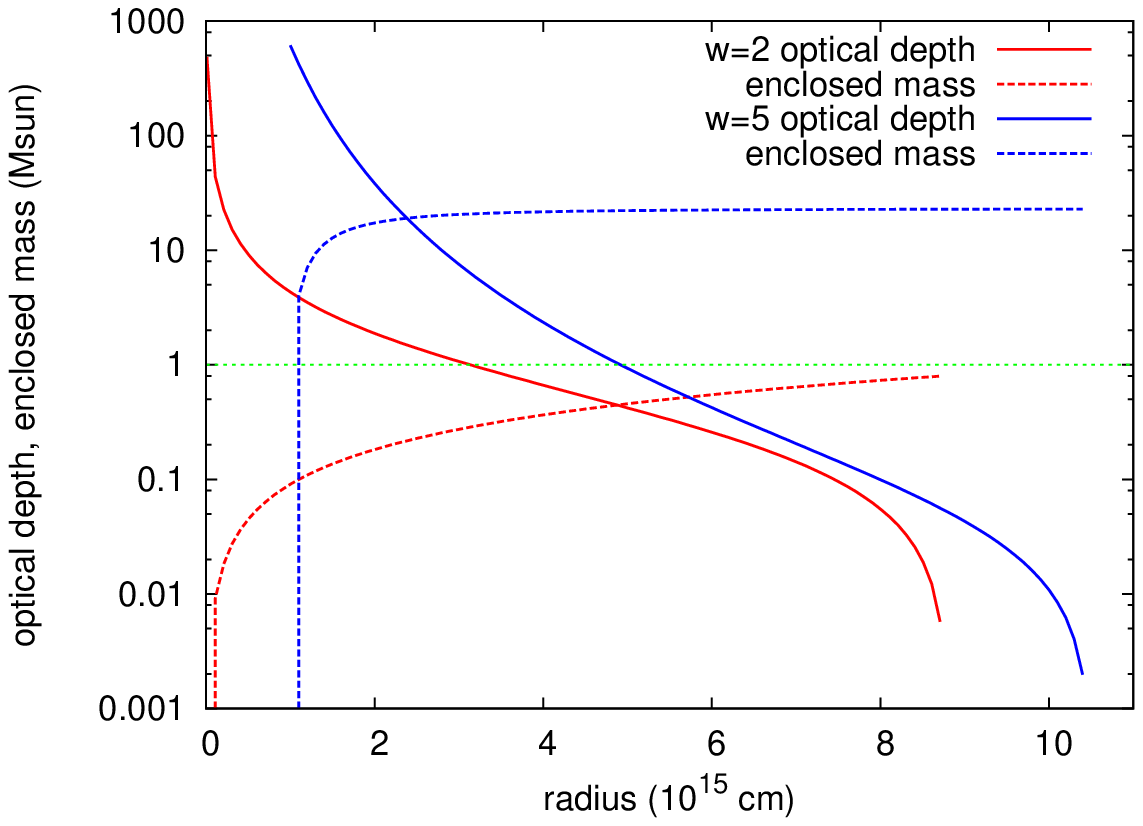}
\includegraphics[width=\columnwidth]{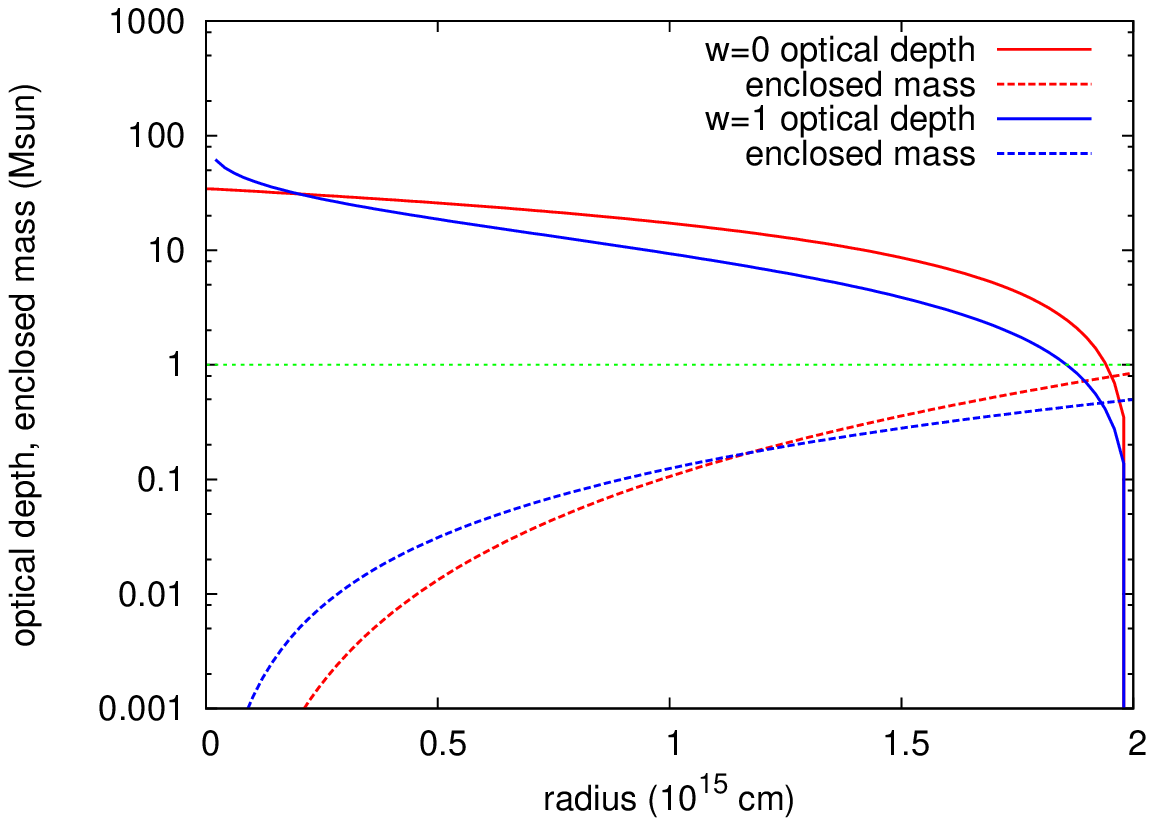}
\caption{
{\it Left:}
Optical depth and enclosed mass distribution of the models $w=2$ and $w=5$ applied for
LSN 2006gy. 
{\it Right:}
Same as the left panel but for the models of LSN 2008es ($w=0,1$).
}\label{fig4}
\end{center}
\end{figure*}

If we adopt the model with $w=2$, for example, $x$ and $R_o$ are
estimated to be $0.0095$ and $8.8\times10^{15}$ cm, respectively.
In this case, shock breakout occurs at
$xR_o\simeq3.2\times 10^{14}$ cm and 
the last scattering surface is $y_1R_o\simeq3.2\times 10^{15}$ cm.
The total wind mass is $M_\mathrm{wind}\simeq0.81~M_\odot$ (c.f. $R_i\ll R_o$)
and is much smaller than the value estimated from the
shell-shocked diffusion LC model \citep[$\sim 10~M_\odot$;][]{smith2007b}.
However, the shell-shocked diffusion model is a too simplified model
and we cannot exclude this model just because of the inconsistency
with it, as is noted in the previous section. 
Alternatively, if we adopt a steeper density gradient $w=5$,
$x$ and $R_o$ are estimated to be $0.17$ and $1.05\times10^{16}$ cm, respectively,
and thus $xR_o\simeq1.8\times 10^{15}~\mathrm{cm}$ and
$y_1R_o\simeq4.9\times 10^{15}$ cm.
$y_1R_o$ is consistent with the blackbody radius at the LC peak
estimated from the observations ($6\times10^{15}$ cm).
If we assume that $R_i\simeq10^{15}~\mathrm{cm}$,
the mass contained in the optically thick region $(R_i<r<y_1R_o)$ is $22~M_\odot$ in our $w=5$ model.
In this case, the mass of the entire wind becomes
$M_\mathrm{wind}\simeq23~M_\odot$.
The left panel of Figure \ref{fig4} is the optical depth
and the enclosed mass distributions.
The existence of the unshocked wind may also account for the
weakness of the X-ray emission of SN 2006gy \citep[e.g.,][]{woosley2007}.

The spectral evolution of SN 2006gy is also consistent with our model.
Lorentzian H Balmer lines seen in the spectra of SN 2006gy 
are presumed to be caused by the optically thick wind
\citep[e.g.,][]{chugai2001,dessart2009}.
For example, for the case $w=5$, $\tau$ become $\simeq 10$ at around
$3\times 10^{15}$ cm (Figure \ref{fig4}).
This is consistent with $\tau\simeq15$ at 36 days which is
estimated from the observed ratio $U$ derived from H$\alpha$
\citep[][]{smith2010}.
In our model,
the Lorentzian line profiles are expected to be observed
before the forward shock wave goes through the optically thick region of
the wind, i.e., before $t_d$, and
thus, the Lorentzian spectra should disappear after the LC peak.
This is also consistent with the spectra of SN 2006gy \citep{smith2010}.

Narrow H$\alpha$ P-Cygni profiles can be created at the optically thin wind above the
last scattering surface of the continuum photons $(y_1R_o<r<R_o)$
because of the larger line opacities.
Whether the narrow H$\alpha$ P-Cygni profiles can be formed or not
depends also on the ionization level of the wind and thus the spectral modeling must be performed
to see whether the narrow H$\alpha$ profiles are actually synthesized in the
unshocked wind in our model.

\subsection{Type IIL LSNe (SN 2008es)}\label{sec3.2}
Here, we apply our model to SN 2008es which is
one of the best observed Type IIL LSNe \citep{miller2009,gezari2009}.
Although SN 2008es does not show any features of the wind in the spectra,
we assume that it is also illuminated by
the interaction based on their brightness, rapid decline of the LC, and blue spectra.
In our model, the lack of the wind features after the LC peak
can be explained by the small difference in $t_d$ and $t_s$ because the
entire wind is shocked by the forward shock just after the LC peak.
In other words, $y_1$ should be close to 1.
This can be achieved by the wind slope with
$w \hspace{0.3em}\raisebox{0.4ex}{$<$}\hspace{-0.75em}\raisebox{-.7ex}{$\sim$}\hspace{0.3em} 1$
for the case of $x<1$ (Figure \ref{fig2}).

The following parameters are estimated from the observations:
\begin{eqnarray}
v_s&\simeq&10,000~\mathrm{km~s^{-1}}, \\
t_d\simeq t_s&\simeq&23~\mathrm{days}.
\end{eqnarray}
If we adopt the model of $w=0$, $y_1$ is close to 1
and $R_o\simeq v_st_s\simeq y_1R_o\simeq 2\times 10^{15}~\mathrm{cm}$.
$y_1R_o$ is consistent with the blackbody radius at the LC peak
estimated from the observations ($3\times10^{15}$ cm).
Assuming $x=0.1$ and $R_i\ll R_o$,
the wind mass is $M_\mathrm{wind}\simeq 0.85~M_\odot$.
$M_\mathrm{wind}$ does not vary so much on $x$ unless it is close to 1.
The model with $w=1$ also gives $y_1\simeq1$ with
$M_\mathrm{wind}\simeq 0.50~M_\odot$.
The right panel of Figure \ref{fig4} shows the optical depth and mass distributions.
\citet{miller2009} estimate
$M_\mathrm{wind}\simeq 5~M_\odot$ based on the shell-shocked diffusion LC model of \citet{smith2007b}.
\citet{gezari2009} obtained $M_\mathrm{wind}\simeq 0.2~M_\odot$ based on
the peak luminosity.
$M_\mathrm{wind}$ of both $w=0$ and $w=1$ models 
are almost consistent with those estimates.

Lack of the Lorentzian H Balmer lines before the LC peak is another
important difference between Type IIL LSN 2008es and Type IIn LSN 2006gy.
For the case of the flat density slope,
the wind optical depth remains to be very large until the forward shock
reaches the wind surface (Figure \ref{fig4}).
Hence, narrow H lines, even if they are emitted from
the dense region of the wind,
are scattered by the dense wind
with a large optical depth for a long while.
Then, $U$ will be very small
until the forward shock reaches the wind surface
and the Lorentzian H lines will be very weak.
Thus, it is likely that the Lorentzian H lines are missed.
Detailed modeling is required to see the actual spectral evolution
expected from the density profile of our model.

\section{Discussion}\label{sec4}
We have shown that the difference in density slopes of
dense winds can make a variety of LSNe
after the shock breakout in the dense wind.
Flat density slopes result in Type IIL LSNe and
steep density slopes result in Type IIn LSNe.
A model with the shock breakout in the dense wind
is recently applied to LSNe by \citet{chevalier2011}.
Their idea for SN 2006gy is basically the same as our suggestion
for Type IIn LSNe: the shock breakout inside the dense wind $(x< 1)$.
However, as they only consider the case $w=2$,
non-Type IIn LSNe are related to
the shock breakout at the surface $(x\simeq1)$ of the dense wind.
In this paper, we show that the shock breakout does not necessarily occur at
the surface to explain non-Type IIn LSNe, especially Type IIL LSNe,
if the progenitor star experiences non-steady mass loss.
Currently, both models can explain Type IIL LSNe.
For the case of the shock breakout inside the dense wind $(x<1)$,
the Lorentzian spectral lines might be able to be observed just before
the LC peak when a suitable optical depth is realized.
The detailed spectral observations near the LC peak can 
distinguish the two scenarios.
Note that we do not exclude the possiblity that steep density slopes
can become Type IIL LSNe. If the density is high enough up to the
surface, steep dense winds can end up with Type IIL LSNe, i.e.,
$y_1$ becomes close to 1 at $x=1$ no matter what the
density decline is. This configuration corresponds to the shock breakout at the
surface and exactly the same as what is suggested by \citet{chevalier2011}.

An important difference between our treatment of the shock breakout in
a dense wind from those of the previous works is
that we adopt Equation (\ref{breakout}) for the shock breakout
condition instead of Equation (\ref{tsb}).
If we use Equation (\ref{tsb}), the differences caused by
the different density slopes are missed.
For example, with Equation (\ref{tsb}), we do not expect to see
narrow spectral lines from the dense wind after the LC peak
of all LSNe with shock breakout in the dense wind 
and they are all expected to be observed as Type IIL SNe.
This is because the forward shock is regarded to
go through the entire wind at the LC peak in Equation (\ref{tsb}).

The different density slope in the dense wind
is naturally expected to be caused by the non-steady mass loss of the progenitor just
before the explosion\footnote{
Note that the flat density distribution of the wind can also
be caused by the steady mass loss of two different evolutionary stages
\citep[e.g.,][]{dwarkadas2011}.
Although the model shown in \citet{dwarkadas2011} is not dense enough to
result in LSNe, 
the flat density slope might result in
Type IIL LSNe if sufficiently high density is achieved.}.
If Type IIL LSNe are actually caused by
the shock breakout in the wind with $t_d/t_s\simeq 1$,
it indicates that non-steady mass loss producing 
the flat dense wind
$(w \hspace{0.3em}\raisebox{0.4ex}{$<$}\hspace{-0.75em}\raisebox{-.7ex}{$\sim$}\hspace{0.3em}1)$
may take place just before the explosions of some massive stars.
In addition, our model for Type IIn LSN 2006gy
prefers $w\neq2$ because $M_\mathrm{wind}$
of the $w=2$ model may be too small to account for the LC of SN 2006gy after
the peak.
This also supports the existence of the non-steady mass loss at the
pre-SN stage of the massive stars.
\citet{dwarkadas2011b} show that the density slope of the wind
estimated from X-ray observations of Type IIn SNe is inconsistent with
$\rho\propto r^{-2}$.
They show that Type IIn SNe do not usually come from the steady wind
with $\rho\propto r^{-2}$.
X-ray luminous Type IIn SNe are presumed to be originated
from relatively dense winds with high mass-loss rates.
Although the wind densities of these SNe are not high enough to be LSNe,
it is highly possible that the dense winds
from higher mass-loss rates also result in flat or steep density slopes.
The presence of the two kinds of slopes can end up with two
different kinds of Type II LSNe.

So far, we just consider a single slope for the dense wind.
One essential difference between Type IIn and Type IIL LSNe
is the existence of the spatially-large optically-thin region in 
the wind of Type IIn LSNe
which can make narrow P-Cygni profiles.
Although we show that large $w$ can make such spatially-large
optically-thin region with the optically thick region inside,
the similar condition can also be achieved by assuming the two
components in the wind, i.e., optically thick (inside) and thin
(outside) regions with any density slopes.
The two-component wind configuration is suggested for, e.g., Type IIn SN
1998S \citep{chugai2001}.
Both models can explain Type IIn LSNe. In either case,
the P-Cygni profiles can be observed not only after but
also before the LC peak. Currently, there are no spectral observations
of Type IIn LSNe before the LC peak
with resolutions sufficient to resolve the narrow
P-Cygni profile and the high resolution spectroscopic
observations before the LC peak
are important to reveal the wind surrounding LSNe.

Our model cannot be simply extended to the spectral evolution of other
kinds of LSNe without H lines.
Especially, Type Ic LSNe with fast LC decline
show, e.g., Si and O lines which are not seen in Types II LSNe
\citep[e.g.,][]{quimby2011,pastorello2010,chomiuk2011}.
Although it is possible that the shock breakout in a dense wind also occurs in
Type Ic LSNe as is suggested by \citet{chevalier2011},
it seems to be difficult to attribute the difference
between Type Ic LSNe and Type II LSNe only to the density slope of the
dense wind.
For example, the composition of the wind is presumed to be
quite different between Type Ic and Type II LSNe.
If the shock breakout in the dense wind is also taking place in Type Ic
LSNe, narrow spectral lines from the materials other than H may be
observed.

While we focus on the origin of Type IIL LSNe in this paper,
the understanding of other Type IIL SNe, i.e., less-luminous Type IIL
SNe, is also lacking.
Currently, there are many models for Type IIn SNe but
only a few models exist for Type IIL SNe \citep[e.g.,][]{blinnikov1993,swartz1991}.
Although the diversity in the wind condition 
may be related to other Type IIL SNe,
there can be other important, but currently ignored,
ingredients for the full understanding of Type IIL SNe.

\section{Conclusions}\label{sec5}
We investigate the effect of the non-steady mass loss on
the shock breakout in the dense wind.
The non-steady mass loss varies the density slope of the wind $(\rho\propto r^{-w})$
and the density slope alters
the ratio of the diffusion timescale in the optically thick wind ($t_d$)
and the shock propagation timescale of the entire wind ($t_s$) after the
shock breakout in the wind.
Both timescales are comparable $(t_d/t_s\simeq 1)$ for
$w\hspace{0.3em}\raisebox{0.4ex}{$<$}\hspace{-0.75em}\raisebox{-.7ex}{$\sim$}\hspace{0.3em}1$
and $t_d/t_s$ becomes smaller as $w$ gets larger.
This is because the last scattering surface of the dense wind
locates farther inside from the wind surface for the wind with the steeper
density gradient (Figure \ref{fig2}).
The difference can only be obtained by the careful treatment of the shock
breakout condition in the dense wind (Section \ref{sec2}; Equation (\ref{breakout})).

If the two timescales are comparable $(t_d/t_s\simeq 1)$,
the forward shock goes through the entire wind just after the LC reaches the peak with the timescale $t_d$.
In this case, no signature on the spectra from the wind is expected to be observed especially
after the LC peak because the entire wind is already shocked after the
LC peak.
On the other hand, if the two timescales are different
($t_d/t_s< 1$),
the shock continues to propagate in the wind after the LC peak
and the unshocked wind remains after the LC peak.
Thus, narrow P-Cygni profiles from the wind are expected to be observed even
after the LC peak.
The former case corresponds to Type IIL LSNe and the latter to Type IIn
LSNe.
The difference in the density slope can also account for the lack of
the Lorentzian emission profiles in Type IIL LSNe.

Our results imply that the luminosity of Type IIL LSNe can be explained in the context of
the shock interaction between SN ejecta and the dense wind even if they do not show the
signature of the wind in their spectra.
We propose that
the difference between Type IIn and Type IIL LSNe can
stem from the density slope of the dense wind which results from the non-steady
mass loss of their progenitors.

\begin{acknowledgments} 
We thank Sergei I. Blinnikov for
the comments on the manuscript and
the continuous discussion regarding to radiation hydrodynamics.
We also thank Keiichi Maeda for discussion.
T.J.M. is supported by the Japan Society for the Promotion of Science Research Fellowship for Young Scientists.
This research is also supported by World Premier International Research
 Center Initiative, MEXT, Japan
and by the Grant-in-Aid for Scientific Research of the
Japan Society for the Promotion of Science (23740157, 23224004).
\end{acknowledgments}

\bibliographystyle{apj}

\end{document}